\documentclass{PoS}

\title{
\vspace{-1cm}
\begin{minipage}{\textwidth}
\begin{flushright}
\normalsize PoS(LAT2009)080\\
\normalsize ITEP-LAT/2009-14
\end{flushright}
\end{minipage}\\[15pt]
Numerical study of chiral magnetic effect in quenched SU(2) lattice gauge theory
\thanks{
This work was partly supported by Grants RFBR
Nos. 08-02-00661-a, and DFG-RFBR 436 RUS, BRFBR
F08D-005, grant for scientific schools No. NSh-679.2008.2,  by the
Russian Federal Agency for Nuclear Power. The calculations were
partially done on the MVS 100K at Moscow Joint Supercomputer
Center. P. V. Buividovich was also supported by the personal grant of the ``Dynasty'' foundation.}}

\ShortTitle{Numerical study of chiral magnetic effect in quenched SU(2) lattice gauge theory}

\author{P.V.~Buividovich~,\\
Joint Institute for Nuclear Research, Dubna, Moscow region, Russia\,\, and \\
Institute for Theoretical and Experimental Physics, B.Cheremushkinskaya 25, Moscow 117259, Russia\\
E-mail: \email{gbuividovich@gmail.com}}

\author{M.N.~Chernodub~,\\
CNRS, Laboratoire de Math\'ematiques et Physique Th\'eorique, F\'ed\'eration Denis Poisson, Universit\'e de Tours, 37200 France\\
Department of Mathematical Physics and Astronomy, University of Gent, Krijgslaan 281, S9, Gent, B-9000 Belgium\, and \\
Institute for Theoretical and Experimental Physics, B.Cheremushkinskaya 25, Moscow 117259, Russia\\
E-mail: \email{maxim.chernodub@lmpt.univ-tours.fr}}

\author{\speaker{E.V.~Luschevskaya} and M.I.~Polikarpov\\
Institute for Theoretical and Experimental Physics, B.Cheremushkinskaya 25, Moscow 117259, Russia\\
E-mail: \email{luschevskaya@itep.ru}\quad
        \email{polykarp@itep.ru}}

\abstract{
A possible experimental observation of the chiral magnetic effect in heavy ion collisions at RHIC was recently
reported by the STAR Collaboration. We study signatures of this effect in SU(2) lattice gluodynamics with the chirally invariant
Dirac operator. We find that at zero temperature the local fluctuations of an electric current of quarks and chirality fluctuations
increase with external Abelian magnetic field. The external magnetic field leads to spatial separation of the
quark's electric charges. The separation increases with the strength of the magnetic field.  As temperature gets higher the dependence
of these quantities on the strength of the magnetic field becomes weaker. In the deconfinement phase the local fluctuations of the chiral density
and of the spatial components of the quarks electric current are large and are almost independent on the external magnetic field. The
local fluctuations of the electric charge density decrease with the strength of the magnetic field in this phase.}

\FullConference{The XXVII International Symposium on Lattice Field Theory - LAT2009\\
         July 26-31 2009\\
         Peking University, Beijing, China}

\begin{document}

\section{Introduction}

Very strong magnetic fields of the hadronic scale can significantly modify the properties of strongly interacting matter:
they change the order of the phase transition from the confinement phase to the quark-gluon plasma, shift the position of
the transition line~\cite{Agasian:2008}, etc. At the Relativistic Heavy Ion Collider (RHIC) in  noncentral
heavy-ion  collisions the strong magnetic field arises due to the relative motion of
the ions and products of the collision~\cite{Kharzeev:McLerran:2008}. The induced magnetic field is perpendicular to the 
reaction plane. At first moments ($\tau~ \sim 1~ \mbox{fm}/\mbox{c}$) of the collision the value of  magnetic field at RHIC may reach
the hadronic scale, $\sqrt{qB}~ \sim ~(10-100~ \mbox{MeV})$ \cite{Kharzeev:McLerran:2008, Skokov:2009}.
Such strong magnetic fields can also be created in future at the ALICE experiment at LHC, at the Facility for Antiproton 
and Ion Research (FAIR) at GSI, and at the Nuclotron Ion Collider fAcility (NICA) in Dubna.

The strong magnetic fields can also enhance the chiral symmetry breaking.
Chiral perturbation theory predicts a linear rise of the chiral condensate $\langle \bar{\psi}\psi\rangle$
with the strength of magnetic field \cite{Shushpanov:1997}, that is in accordance with our lattice results \cite{Buividovich:2008}. In the
AdS/QCD approach \cite{Zaykin:2008}, in Nambu-Jona-Lasinio model
\cite{Klevansky:1989} and in sigma models
\cite{Goyal:2000} the value of $\langle \bar{\psi}\psi\rangle$ rises
quadratically with the external magnetic field.

Strong magnetic fields lead also to the chiral magnetization of QCD vacuum at zero and finite
temperature. This effect has a paramagnetic nature because the quarks
are 1/2-spin particles and the external magnetic field leads to
polarization of magnetic moments of quarks in the direction of the
magnetic field. The magnetization of the QCD vacuum is essential for the properties
of the nucleon magnetic moments \cite{Ioffe:1984} and other nonperturbative features of hadrons~\cite{Pire:2009:Rohrwild:2007}.
Our lattice simulations~\cite{Buividovich:Chernodub:2009} demonstrate
at the relatively weak magnetic fields the chiral magnetization rises
linearly in accordance with Ref.~\cite{Ioffe:1984}.
At larger values of the magnetic fields the magnetization depends nonlinearly on the magnetic field~\cite{ref:Thomas}.

We also observed an anomalous quark magnetization in intense magnetic fields
which appears due to the topological fluctuations of the QCD
vacuum~\cite{Buividovich:Luschevskaya:2009}. The essence of this effect is
that quark gets a local electric dipole moment in addition to the magnetic moment.
Both magnetic and anomalous electric moments are parallel to the axis of the
external magnetic field. The generation of the anomalous electric dipole moment is a spin analogue of the chiral magnetic effect (CME).

CME is the generation of a local electric current in the direction of the external magnetic field in topologically nontrivial
configurations of the gauge fields~\cite{Kharzeev:McLerran:2008,Fukushima:2008}. If we
consider $u$ and $d$-quarks as massless particles then the
right-handed quarks should move in the direction of the magnetic
field and the left-handed quarks should move in the opposite
direction  because in the external field magnetic moments of
quarks are parallel to the direction of the field.
Nonzero topological charge of gauge fields
leads to a local imbalance
between left-handed and right-handed quarks, which, in turn, gives rise to a
nonzero net electric current along the axis of the magnetic field.
An evidence of the CME was observed by the STAR Collaboration at RHIC as an
non-statistical asymmetry of negatively and positively charged particles emitted
at different sides of the reaction plane
of noncentral heavy-ion collisions~\cite{Voloshin:2008,Caines:2009}. The observed effect indicates
the presence of the chirality fluctuations
in heavy-ion collisions.
Below we report the evidence of the CME in the lattice gauge theory following our original study~\cite{ref:original}.

\section{Details of calculations}

We generate statistically independent SU(2) gauge field configurations with the tadpole improved Symanzik
action \cite{Bornyakov:Luschevskaya:2008}. Using the chirally
invariant Dirac operator \cite{Neuberger:1997},
we solve the Dirac equation $D\psi_k = \lambda_k \psi_k$ numerically
and determine the corresponding eigenfunctions $\psi_k$ and eigenvalues $\lambda_k$.
Here $D=\gamma^{\mu}(\partial_{\mu}-iA_{\mu})$ is the massless Dirac operator in the gauge field $A_{\mu}$.
The uniform magnetic field is introduced into this operator as described in Ref.~\cite{Buividovich:2008}.

We performed the zero-temperature simulations
on $14^4$ lattice with the lattice spacing $a = 0.103~\mbox{fm}$
and on $16^4$ lattice
with the lattice spacings
and $a = 0.103~\mbox{fm}$ and $a = 0.089~\mbox{fm}$.
At finite temperature we used $16^3 \times 6$ lattices with the
spacings $a = 0.128~\mbox{fm}$ ($T = 256~\mbox{MeV} = 0.82~T_c$) and
$a = 0.095~\mbox{fm}$ ($T = 350~\mbox{MeV} = 1.12~ T_c$). The
critical temperature of the phase transition in $SU(2)$  gauge
theory is $T_c = 313.(3)~ \mbox{MeV}$ \cite{Bornyakov:Ilgenfritz:2007}.
The values of the magnetic field are quantized
due to
the periodic boundary conditions imposed in
finite lattice volume.
In our simulations a  minimal nonzero value of  magnetic field is $qB_{min}=(348~\mbox{MeV})^2$.

\section{Fluctuations of electric charge density}

In order to study the
simplest
signatures of the chiral magnetic effect we
first
explore the fluctuations of the electromagnetic current
\begin{equation}
j_{\mu}(x) = \bar{\psi}(x) \gamma_{\mu} \psi(x).
\label{current}
\end{equation}
The zeroth component of (\ref{current}) corresponds to a local charge density.
We remove the diverging ultraviolet contributions by subtracting the charge density at zero magnetic field $B$
from the charge density calculated at nonzero $B$.
\begin{figure}
\center{\vspace{-2.7cm}
\begin{tabular}[h]{cc}
\hspace{1cm}
\includegraphics[scale=0.29, angle=0]{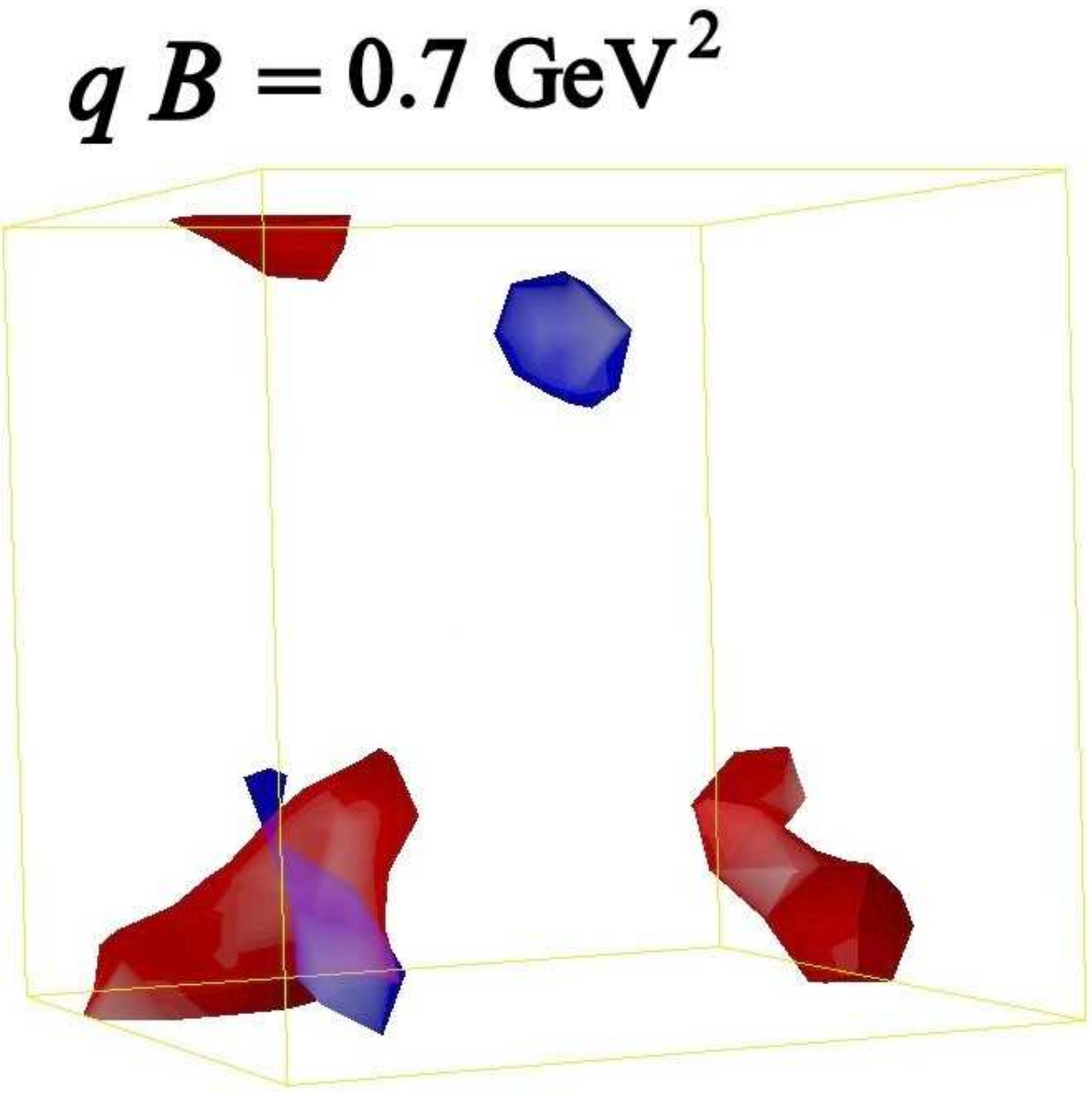} \hspace{0.5cm} &
\includegraphics[scale=0.29, angle=0]{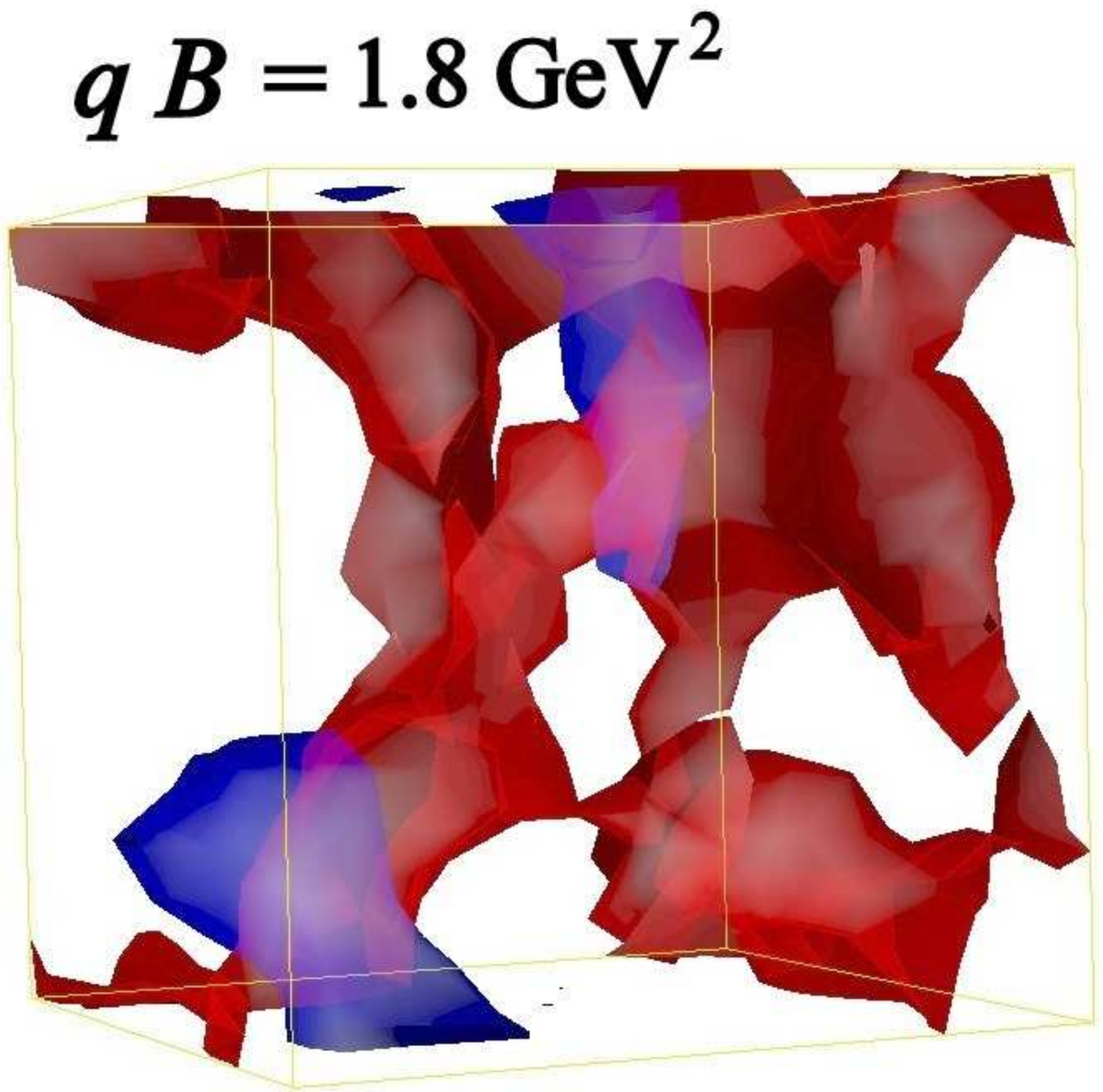}\\
\end{tabular}}\\
\caption{The excess of the electric charge density due to the external magnetic field $qB = 0.7~\mbox{GeV}^2$ (left) and
$qB=1.81~\mbox{GeV}^2$ (right) on $14^4$ lattice corresponding to the physical volume of
$(1.44~\mbox{fm})^4$. The red regions corresponds
to the excess of the positive charge density, while the blue color marks the
excess of the negative charge density.} \label{fig:fig1}
\end{figure}
The result of the subtraction is a nonperturbative infrared quantity which is plotted in
Fig.~\ref{fig:fig1}. Here we visualize the level surfaces of two typical
charge density distributions in a fixed time slice of the
zero-temperature gauge field configuration for two values of the
external magnetic field $B$. The magnetic field is directed from the
bottom to the top.

We see in Fig.~\ref{fig:fig1} that the local charge density grows with the external magnetic field.
The regions with larger charge density are extended along the direction of the magnetic field
because at a sufficiently strong field the quarks occupy the lowest Landau level. In
the absence of the gluon background
the trajectories of the quarks should be elongated along the direction of the magnetic field.
The gluon field distorts the motion of quarks, so that the trajectories of quarks diffuse.

\section{Fluctuations of chirality}

The other quantity which characterizes the chiral magnetic effect is the local chirality:
\begin{equation}
\rho_5(x)=\bar{\psi}(x) \gamma_5 \psi(x).
\label{chirality}
\end{equation}
The average value of (\ref{chirality}) equals to zero, 
therefore we measure
the squared chirality:
\begin{equation}
\langle \rho_5^2 \rangle_{IR}(B,T)=\frac{1}{V}\int_V d^4x \langle \rho_5(x)^2 \rangle_{B,T} - \frac{1}{V}\int_V d^4x \langle \rho_5(x)^2 \rangle_{B=0,T=0},
\label{squared:chirality}
\end{equation}
where $V$ is the lattice volume. The subtraction of the $B=0$ and $T=0$ quantity is performed in order to regularize 
the total result in the ultraviolet region.

\begin{figure}
\center{
\begin{tabular}[h]{cc}
\hspace{-1cm}
 \includegraphics[scale=0.29, angle=0]{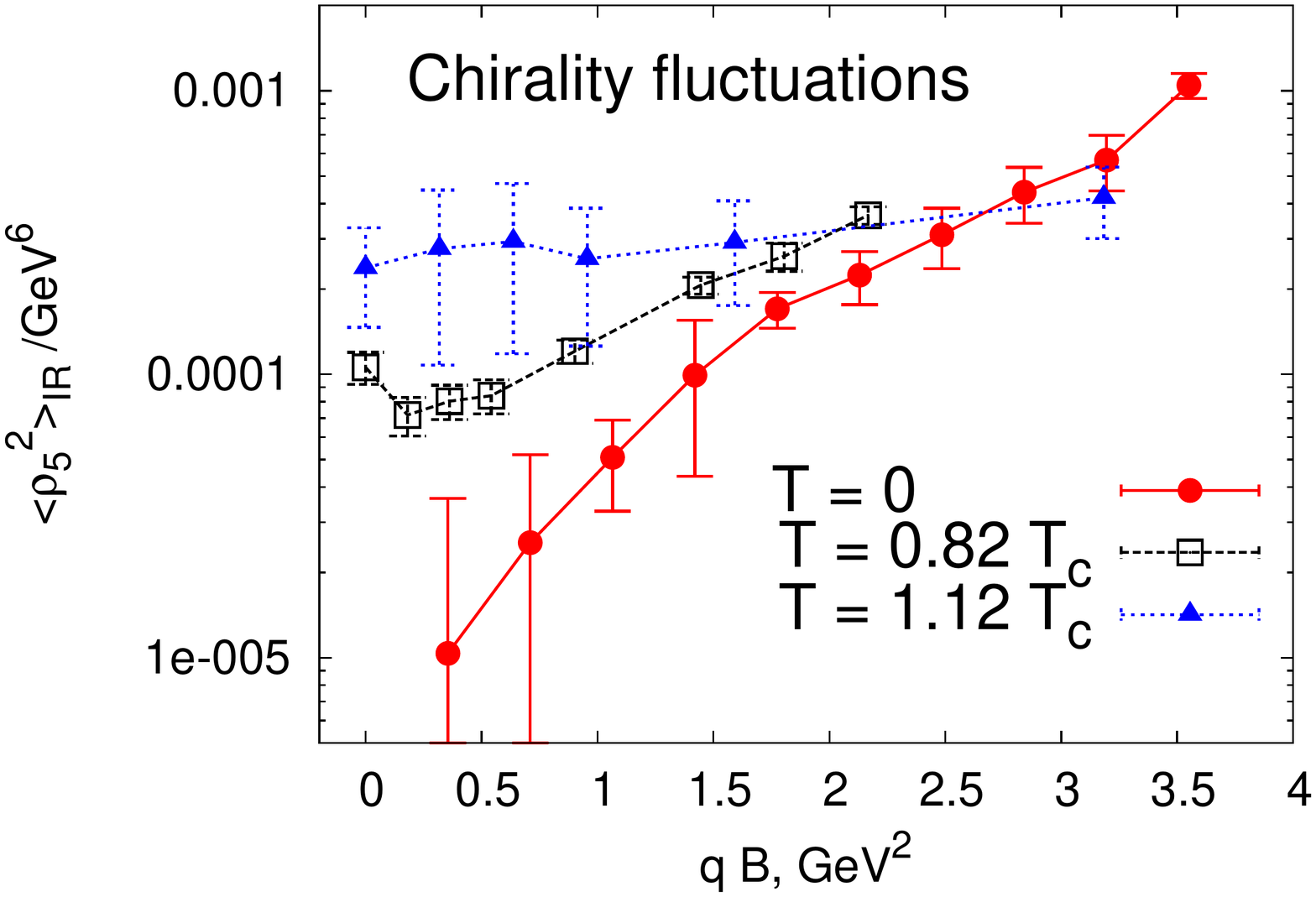}\hspace{-0.7cm}
&
\includegraphics[scale=0.29, angle=0]{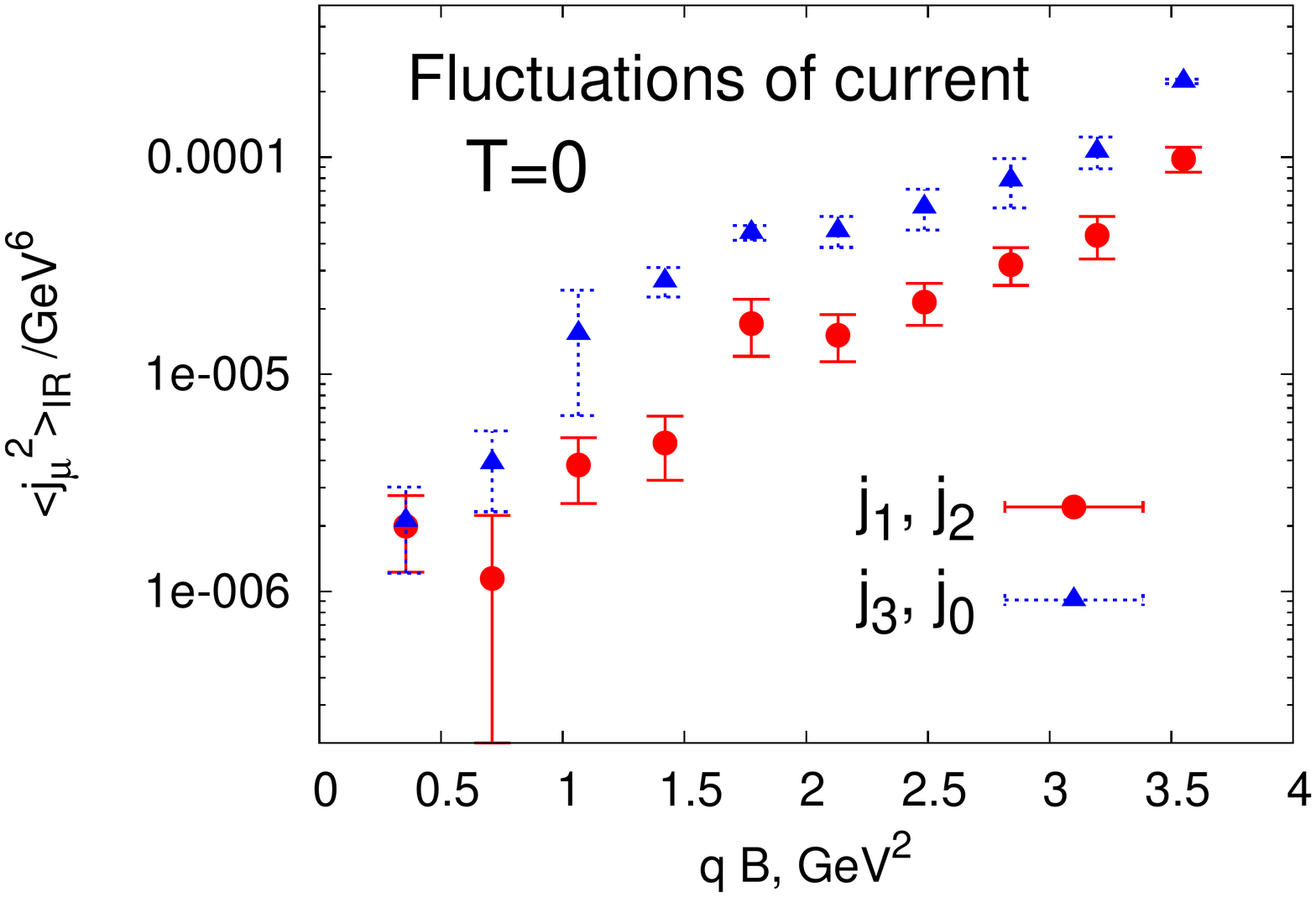}\\
\end{tabular}}\\
\caption{The expectation values of the squared chirality  (left) and  of the squares of the longitudinal ($j_0$ and $j_3$) and the transverse  ($j_1$ and $j_2$) electric currents (right)  vs the external magnetic field at three temperatures.}
 \label{fig:fig2}
\end{figure}

Fig.~\ref{fig:fig2} (left) shows the squared chirality at different temperatures~$T$.
At zero temperature the chirality fluctuations increase quickly with the value of $B$. In the confinement phase at nonzero temperature
($T=0.82~T_c$) the grows rate becomes smaller and in the deconfinement phase ($T=1.12~T_c$) the squared chirality is
almost independent on the value of the magnetic field.

The value of the chirality fluctuations at zero temperature and sufficiently large magnetic
fields ($qB \sim 1 \mbox{GeV}^2$) is close to the value of $\langle \rho_5^2 \rangle$ in the
deconfinement phase ($T=1.12~T_c$) at zero magnetic field.
Thus, we can conclude that strong chirality fluctuations and hence the chiral magnetic effect
could be also observed in a cold nuclear matter exposed to
the strong magnetic fields. However, in this case the experimental signatures 
of the CME can be quite different from those in heavy-ion collisions. 

\section{Fluctuations of electromagnetic current}

We study the fluctuations of the electric current:
\begin{equation}
\langle j_{\mu}^2 \rangle_{IR}(B,T)=\frac{1}{V}\int_V d^4x \langle j^2_{\mu}(x)\rangle_{{B,T}} - \frac{1}{V}\int_V d^4x \langle j^2_{\mu}(x)\rangle_{{B=0,T=0}}.
\label{current:fluc}
\end{equation}
We checked on the lattice, that all components of the electric current (\ref{current}) are on average
equal to zero within statistical error bars. The component of the current along the direction of the magnetic field
is called longitudinal. Transverse components are perpendicular to
the direction of the field.
At $T=0$ the magnetic field in the $\mu=3$ direction breaks the
rotational symmetry, so that we have: $\langle j_1^2 \rangle= \langle
j_2^2  \rangle$ and $\langle j_0^2 \rangle=\langle j_3^2 \rangle$.

In Fig.~\ref{fig:fig2} (right) we show all components of the electric field.
All of them grow with the strength of the field.
The longitudinal component of the electric current and the electric charge
fluctuate stronger than the transverse ones. The fluctuations of
the spatially transverse components grow with the field because the transverse momentum
of a quark occupying the lowest Landau level increases with the strength of the magnetic field.

At $T\neq0$ the rotational symmetry is broken by the temperature and
by the external magnetic field, so that the longitudinal and zero components
are no more equivalent, $\langle j_0^2 \rangle\neq\langle j_3^2 \rangle$,
while the spatially transverse components are still degenerate, $\langle
j_1^2 \rangle= \langle j_2^2  \rangle$.

\begin{figure}
 \center{
\begin{tabular}[h]{cc}
\hspace{-1cm}
 \includegraphics[scale=0.29, angle=0]{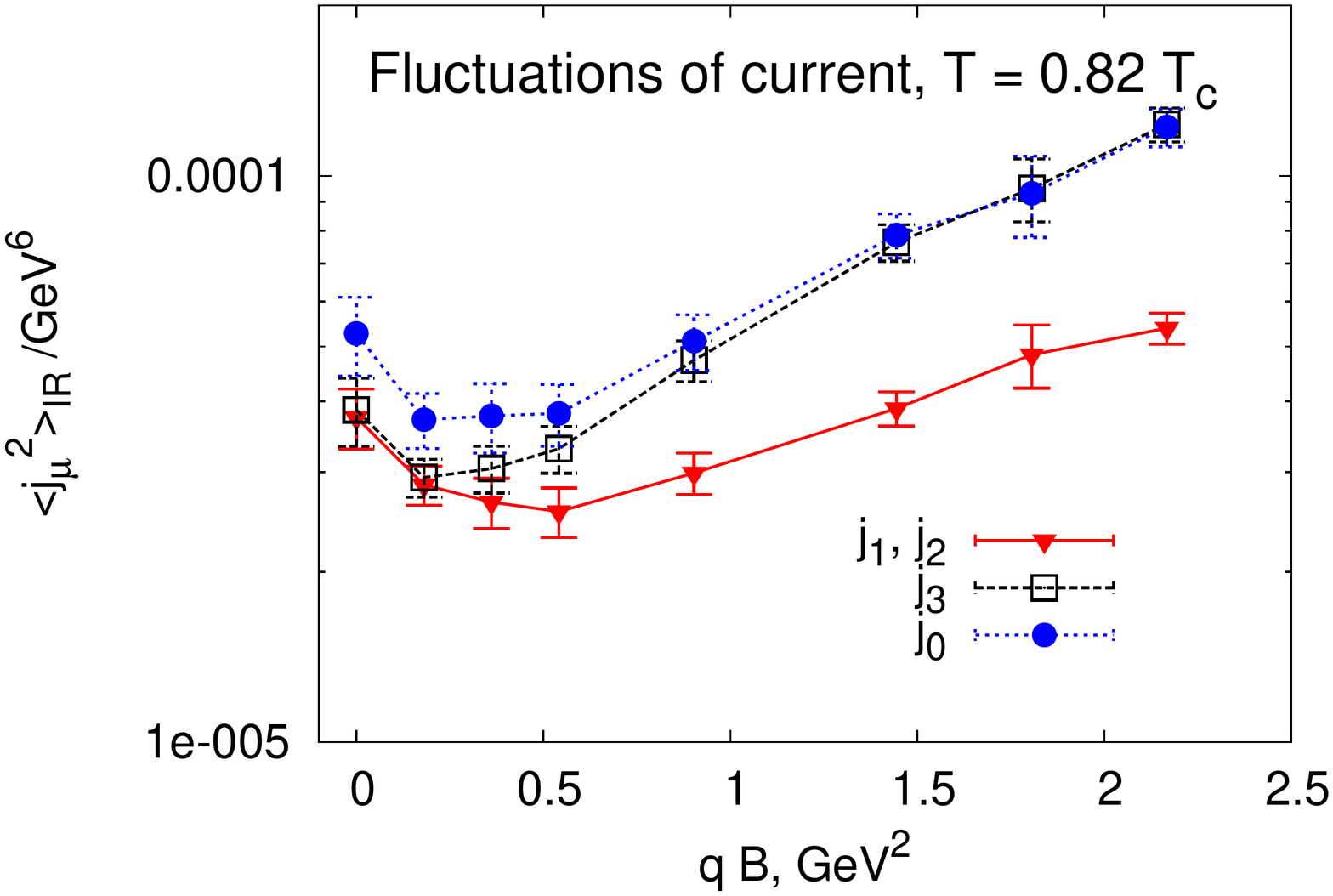} \hspace{-0.7cm}
&
\includegraphics[scale=0.29, angle=0]{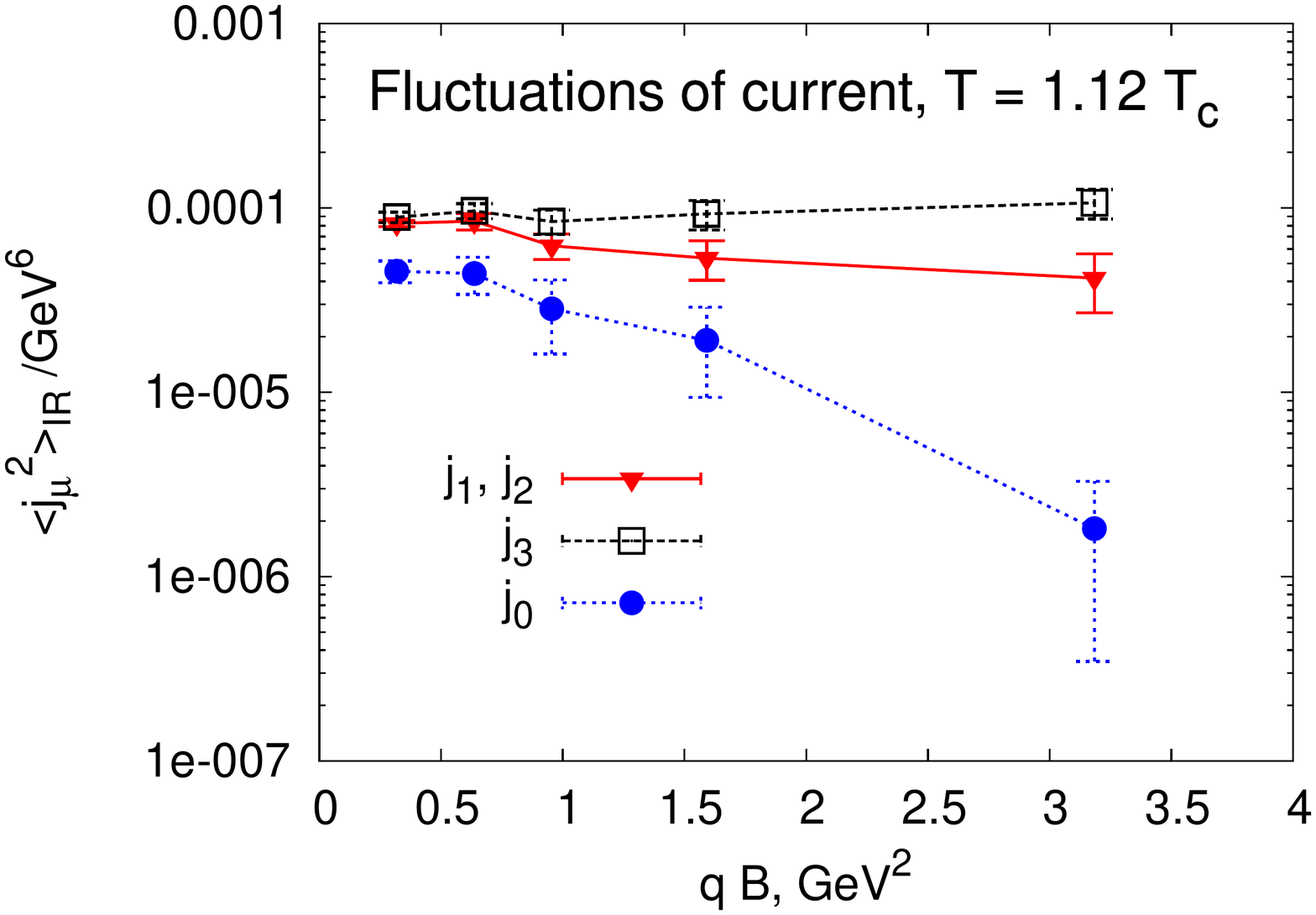}\\
\end{tabular}}\\
\caption{The squares of the longitudinal and transverse components of the
electric currents vs the external magnetic field in the confinement phase (left) and in the deconfinement phase (right).}
\label{fig:fig3}
\end{figure}

In Fig.~\ref{fig:fig3} (left) we show the fluctuations of the electric current in the confinement phase.
At weak fields all components decrease slightly (the decrease is, however, within the statistical errors).
Then the fluctuations start to increase with the value of the
magnetic field. The components $\langle j^2_0 \rangle$ and $\langle j^2_3 \rangle$
are larger than $\langle j^2_1 \rangle$ and $\langle j^2_2 \rangle$.
From Fig.~\ref{fig:fig3} (right) we see, that in the deconfinement
phase the components
$\langle j^2_1 \rangle$, $\langle j^2_2 \rangle$ and $\langle j^2_3 \rangle$
are almost independent on the external magnetic field $B$, while the charge fluctuation
$\langle j^2_0 \rangle$ is a decreasing
function of $B$. This observation can be explained by the Debye screening in the deconfinement phase.

\section{Chirality-current correlations}

We observed that the $\langle \rho_5 j_{\mu} \rangle=0$. This fact is natural due to the $CP$-oddness of this quantity. Next,
we study the correlation between the squares of the chirality and the induced current:
\begin{equation}
c(\rho_5^2,j^2_{\mu})=\frac{\langle \rho^2_5 j^2_{\mu}\rangle - \langle \rho^2_5 \rangle \langle j^2_{\mu}\rangle}{\langle \rho^2_5 \rangle \langle j^2_{\mu}\rangle}\,.
\label{chir:curr:corr}
\end{equation}
\begin{figure}
 \center{
\begin{tabular}[h]{cc}
\hspace{-1cm}
 \includegraphics[scale=0.29, angle=0]{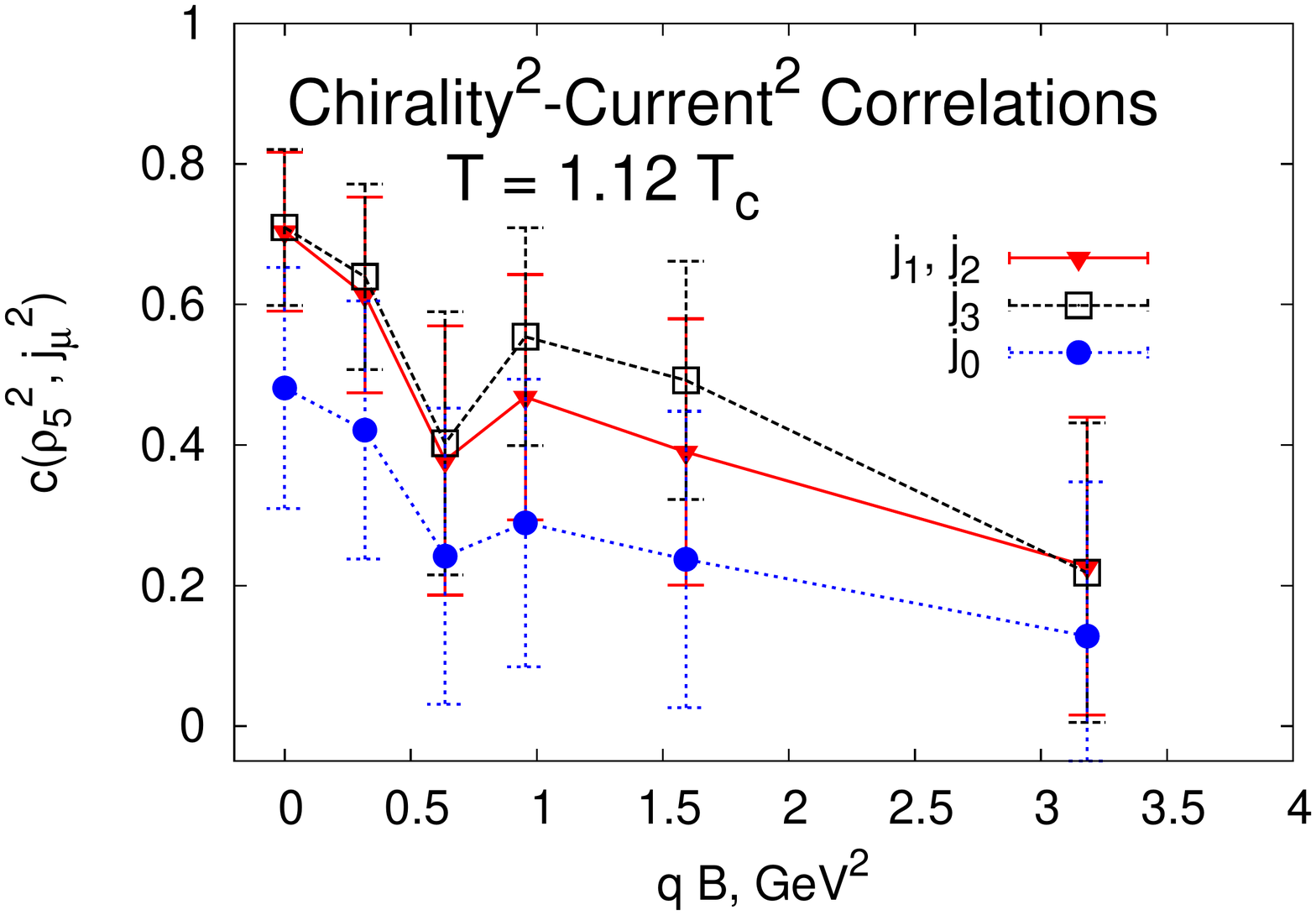}  \hspace{-0.7cm} &
\includegraphics[scale=0.29, angle=0]{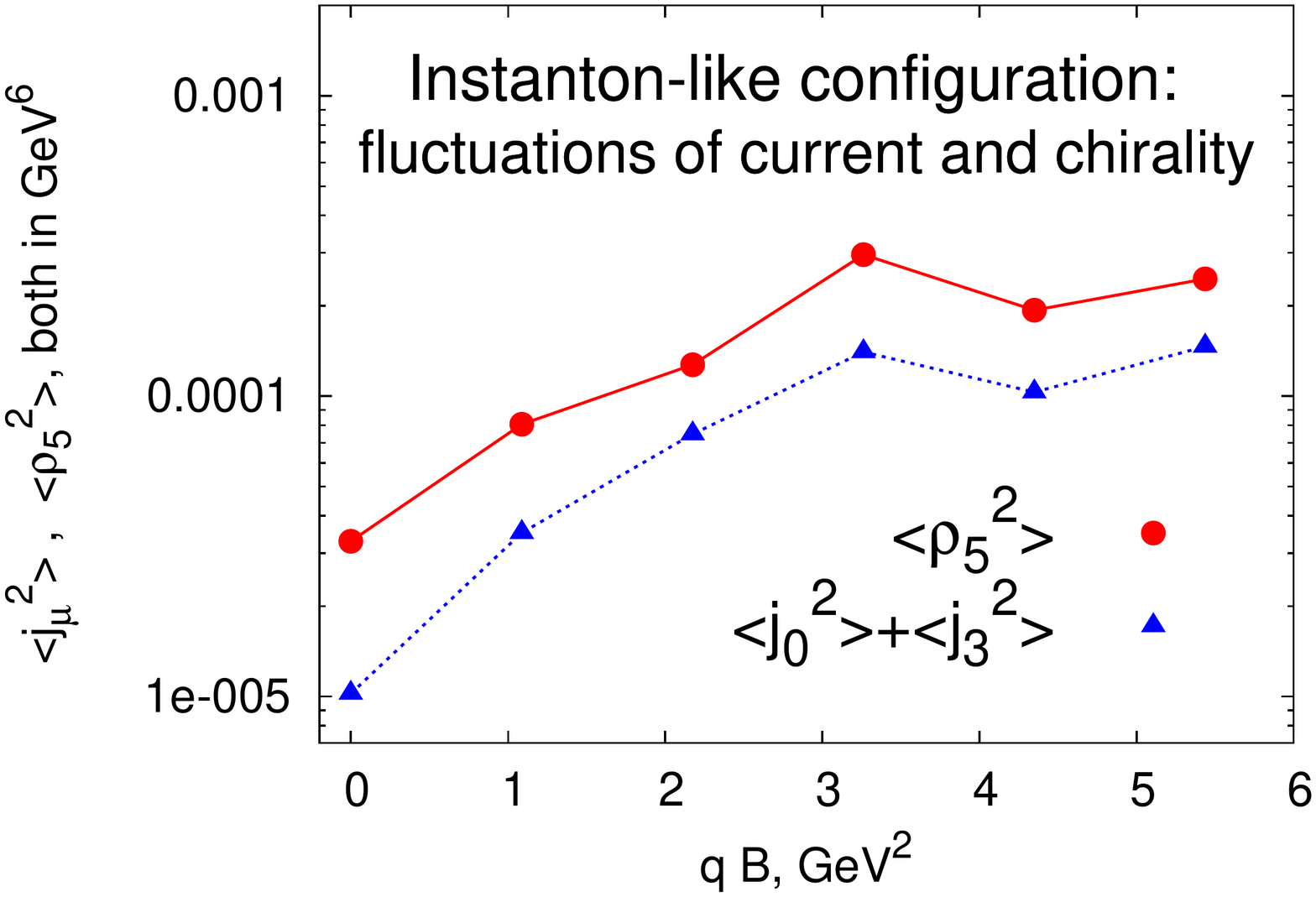}\\
\end{tabular}}\\
\caption{Left: the correlation function of
$\rho^2_5(x)$ and $j^2_{\mu}(x)$,
Eq.~(6.1).
Right: The values of $\langle\rho^2_5 \rangle$  and  $\langle j^2_{\mu}\rangle$ at an instanton-like configuration with $Q=1$.
}
\label{fig:fig4}
\end{figure}
We found that in the confinement phase the value of (\ref{chir:curr:corr}) vanishes for all components
of the electric currents. In the deconfinement phase the correlator  is decreasing function of the field [Fig.~\ref{fig:fig4} (left)].
Therefore, the enhancement of the fluctuations of the electric
current in the direction of the field is not locally correlated with the chirality fluctuations.

\section{Instanton-like configuration}

We have also studied signatures of the chiral magnetic effect at an
instanton-like gauge field configuration with unit topological charge,
$Q=1$. In Fig.~\ref{fig:fig4} (right) we show the fluctuations of the
local chirality and the fluctuations of the squared longitudinal electric
current in the instanton background. We observed the growth of the chiral
density with the strength of the applied external magnetic field. The
grows is caused by a difference between the particles with positive and
negative chiralities in the $Q=1$ background.
The net longitudinal electric current, $j^2_{\parallel}=j^2_0+j^2_3$,
increases with the strength of the field as well. This is
the signature of the chiral magnetic effect.

\section{Conclusion}

We found the signatures of the chiral magnetic effect in lattice
SU(2) gluodynamics in a background of the strong external magnetic field.
This effect arises due to imbalance between left- and right-handed quarks (the local chirality).
The average chirality in a large enough volume
is equal to zero, but the local fluctuations of the chiral density can be sufficiently strong.
We found that the fluctuations of the chirality at zero temperature increase with the strength
of the magnetic field. As the temperature increases the fluctuations become weaker dependent on
the strength of the field $B$. In the deconfinement phase the local chirality fluctuation are almost independent on $B$.

We found that in the confinement phase all components of the electric current of quarks
are growing in intense magnetic fields. The longitudinal (with respect to the axis of the magnetic field)
currents fluctuate stronger than the transverse ones. In the deconfinement phase the thermal fluctuations
lead to the increase of the current fluctuations while all spatial components of the current are insensitive to the
value of the magnetic field. The charge fluctuations decrease as the field gets stronger. The fluctuations of the electric
current and chirality are almost independent.

Finally, we observed the chiral magnetic effect at the instanton-like configuration.

\end{document}